\documentclass[%
 reprint,
 superscriptaddress,
nofootinbib,
 amsmath,amssymb,
 aps,
 prd,
]{revtex4-2}

\usepackage{graphicx}
\usepackage{dcolumn}
\usepackage{bm}
\usepackage{amsmath}
\usepackage{amssymb}
\usepackage{amsfonts}
\usepackage{mathtools}
\usepackage{url}
\usepackage[dvipsnames]{xcolor}
\usepackage[colorlinks]{hyperref}
\usepackage{float}
\usepackage[inline]{enumitem}
\usepackage[]{subcaption}
\usepackage{appendix}

\definecolor{burgundy}{rgb}{0.6, 0.0, 0.0}
\definecolor{persianblue}{rgb}{0.11, 0.22, 0.73}
\definecolor{darkblue}{rgb}{0.0,0.0,0.4}
\hypersetup{colorlinks, citecolor=darkblue, linkcolor=persianblue, urlcolor=persianblue}

\def\apj{ApJ}

\begin{document}

\title{Diffuse $\gamma$-ray emission in Cygnus X: Comments to Yan\&Pavaskar https://arxiv.org/abs/2211.17057}

\author{Ottavio Fornieri}
\affiliation{Gran Sasso Science Institute, Viale F. Crispi 7, 67100 L’Aquila, Italy}
\affiliation{INFN - Laboratori Nazionali del Gran Sasso (LNGS), Via G. Acitelli 22, 67100 Assergi (AQ), Italy}

\author{Heshou Zhang}
\affiliation{INAF - Osservatorio Astronomico di Brera, Via E. Bianchi 46, 23807 Merate (LC), Italy.}

\begin{abstract}
\vspace{0.2cm}
\begin{center}
    \textbf{Abstract}
    \vspace{-1.cm}
\end{center}
It has been pointed out in https://arxiv.org/abs/2211.17057~\citep{Yan:2022mma} that our recent (published) paper~\citep{PhysRevD.106.103015} might be revised, due to an incorrect evaluation of the diffusion coefficients, $D(E)$, employed in the calculations. Unfortunately, there is no indication as to where the incorrectness might be. Here we offer the opportunity to be more specific, by providing the community with the whole description of the equations involved in the calculation of $D(E)$, which is missing in the {\tt arXiv} note~\citep{Yan:2022mma}. In this context, we mention that no \textit{ad hoc} parameterisation has been used in our paper. Furthermore, any assumption on the injection mechanisms is explicitly described in the paper as an input factor and is obviously part of the modelisation procedure, hence the final outcome is subject to it. Finally, we discuss in a more broad context what, in this calculation of the diffusion coefficient, we believe is the key message.
\end{abstract}

\maketitle

\section{Introduction}
To compute the diffusion coefficient that cosmic rays (CRs) feel, $D(E)$, as a function of their energy, it is necessary to estimate the interaction between the particle --- represented by its Larmor radius $r_L$ --- and the turbulent waves, which act as scattering centers for the CRs. These waves, depending on the environment where they are produced, can propagate undisturbed --- therefore contributing to the diffusion coefficient of the particles --- or can be damped. This is formalized in the following equation, to compute the pitch-angle scattering of the CRs, $D_{\mu\mu}$, where the integral bounds up to a maximum wavenumber, $k_{\max}$, account for the damping of the fluctuations at a certain scale~\citep{1969ApJ...156..445K, 1975RvGSP..13..547V}:
\begin{equation}\label{eq:pitch-angle_diffusion_general}
\begin{aligned}
    D_{\mu \mu} &= \Omega^2 (1 - \mu^2) \int d^3 \bm{k} \sum^{+\infty}_{n=-\infty} R_n (k_{\parallel} v_{\parallel} - \omega + n \Omega) \times \\
    &\times \left[ \frac{n^2 J_n^2(z)}{z^2}I^{\rm A} (\bm{k}) + \frac{k^2_{\parallel}}{k^2} J'^{2}_{n}(z) I^{\rm M} (\bm{k}) \right],
\end{aligned}
\end{equation}
where $\Omega=q B_0 /m \gamma c$ is the particle's Larmor frequency; $\bm{k}$ is the wave-vector of the turbulent fluctuations; $k_{\parallel} \equiv |\bm{k}|\cos{\alpha_{w}}$ is its field-aligned component ($\alpha_{w}$ being the angle between the wave vector and the direction of the background magnetic field, $B_0$); $\omega=\omega(\bm{k})$ the associated fluctuations' frequency. The details of the calculation are described in the Appendix of \citet{Fornieri:2020wrr}.

Several damping mechanisms have been considered in the literature. Given the typical environmental parameters considered in our Galaxy, only the \textit{collisionless}~\citep{ginzburg1970propagation} and the \textit{collisional} (viscous)~\citep{1965RvPP....1..205B} dampings have been isolated and are at play~\citep{2006ApJ...644..603P, Fornieri:2020wrr}.

The parameters of the plasma involved in the calculation of the scale at which turbulent fluctuations are damped (in wavenumber $k_{\mathrm{max}}$, normalized to the turbulence injection scale, $k_{\mathrm{max}}L_{\mathrm{inj}}$) --- and therefore cannot act as CR scattering centers --- depend on physical quantities characteristic of the Cygnus-X environment. In particular, we use the following values, as listed in our paper:
\begin{equation}
\begin{aligned}
    &T_{\mathrm{Cyg}} = 5 \cdot 10^3 \, \mathrm{K} \\
    &n_{\mathrm{Cyg}} = 10^{-1} \, \mathrm{cm^{-3}} \\
    &B_{\mathrm{Cyg}} = 10 \, \mu \mathrm{G} \\
    &L_{\mathrm{inj}} = 10 \, \mathrm{pc}.
\end{aligned}
\end{equation}

To compute the truncation scale of the fluctuations, we equate the cascading timescale, for the fast magnetosonic modes, with the (specific) damping rate~\citep{Yan:2007uc}. In particular, provided that for the typical \textit{Interstellar Medium} (ISM) conditions $\beta \ll 1$ everywhere, we have:
\begin{equation}\label{eq:truncation_scale}
\begin{aligned}
    &k_{\mathrm{max}} L_{\mathrm{inj}} = 
    x_c \, (1 - \xi^2)^{-2/3}, \qquad \beta \ll 1 \\[10pt]
    &k_{\mathrm{max}} L_{\mathrm{inj}} = \frac{4 \, M_{\mathrm{A}}^4 \, \frac{m_p}{m_e} \, \xi^2}{\pi \beta \, (1 - \xi^2 )^2} \cdot \exp \left( \frac{2}{\beta \, \frac{m_p}{m_e} \, \xi^2} \right),
\end{aligned}
\end{equation}
where $\xi\equiv \cos \alpha_w$. The equations show the truncation scales for the viscous and collisionless dampings, respectively, while the other quantities are discussed below. From here on, $L = L_{\mathrm{inj}}$ interchangeably, to avoid heavy notation.

The plasma parameters for the \textit{viscous} case can be rearranged as follows:
\begin{widetext}
\begin{equation}\label{eq:viscous_parameters}
\begin{aligned}
    &x_c = \left( \frac{6 \rho \delta u^2 L_\mathrm{inj}}{\eta_0 v_A} \right)^{2/3} = \left( \frac{6 \rho M_A^2 L_\mathrm{inj} \, v_A}{\eta_0} \right)^{2/3}, \qquad M_A \equiv \frac{\delta u}{v_A}\bigg\vert_{\mathrm{inj}} \\[10pt]
    &-\hspace{0.2cm}\rho = n_{\mathrm{Cyg}} \cdot m_p \\
    &-\hspace{0.2cm}\eta_0 = 6 \cdot 10^3 \left( \frac{37}{\ln \Lambda} \right) \left( \frac{T_{\mathrm{Cyg}}}{10^8 \, \mathrm{K}} \right)^{5/2} \, \mathrm{\frac{g}{cm \cdot s}} \\
    &\hspace{0.55cm}\ln \Lambda \equiv \ln \left( \frac{b_{\mathrm{max}}}{b_{\mathrm{min}}} \right) \simeq \ln \left( \frac{\lambda_D}{b_{\mathrm{min}}} \right), \;\; b_{\mathrm{min}} \approx 8.63 \cdot 10^{-8} \left( \frac{10^4 \, \mathrm{K}}{T_{\mathrm{Cyg}}} \right) \, \mathrm{cm} \\
    &-\hspace{0.2cm}\lambda_D \approx 0.95 \cdot 10^5 \left( \frac{T_{\mathrm{Cyg}}}{10^6 \, \mathrm{K}} \right)^{1/2} \left( \frac{10^{-3} \, \mathrm{cm^{-3}}}{n_{\mathrm{Cyg}}} \right)^{1/2} \, \mathrm{cm}
\end{aligned}
\qquad \qquad \qquad
\begin{aligned}
    &\\[25pt]
    &\textrm{density of the medium} \\[10pt]
    &\textrm{Spitzer 1962~\citep{1962pfig.book.....S, 1965RvPP....1..205B}} \\[16pt]
    &\textrm{Coulomb logarithm}\\[18pt]
    &\textrm{Debye lenght}.
\end{aligned}
\end{equation}
\end{widetext}

The details of these equations can be found in \citet{1962pfig.book.....S, 1965RvPP....1..205B, ginzburg1970propagation}.

The plasma parameters for the \textit{collisionless} case are:
\begin{equation}
\begin{aligned}
    &\beta \equiv \frac{P_\mathrm{th}}{P_{B}} = 3.3 \cdot \left( \frac{3 \, \mu\mathrm{G}}{B_{\mathrm{Cyg}}} \right)^2 \left( \frac{n_{\mathrm{Cyg}}}{1 \, \mathrm{cm^{-3}}} \right) \left( \frac{T_{\mathrm{Cyg}}}{10^4 \, \mathrm{K}} \right) \\[6pt]
    &v_A = 6.27 \cdot 10^5 \left( \frac{B_{\mathrm{Cyg}}}{3 \, \mu \mathrm{G}} \right) \left( \frac{1 \, \mathrm{cm^{-3}}}{n_{\mathrm{Cyg}}} \right)^{1/2} \, \mathrm{cm \cdot s^{-1}}
\end{aligned}
\end{equation}
where $\beta$ is the ratio between the thermal pressure and the magnetic pressure in the ISM, and $v_A$ is the usual Alfv\'en speed.

These parameterisations have been used to compute the diffusion coefficients in the paper in question~\citep{PhysRevD.106.103015} and in \citet{Fornieri:2020wrr}, and for both of them have been checked multiple times by multiple persons, other than the reviewers, obviously. In any case, this is what should be addressed to find any (certainly possible) problems in the calculation, with an \textit{independent} estimation. Again, we would like to stress that no \textit{ad hoc} parameterisation has been involved, unlike what claimed in the {\tt arXiv} note~\citep{Yan:2022mma}. 

\begin{figure*}[t!]
\centering
    \begin{subfigure}{0.49\textwidth}
        \centering
        \includegraphics[width=0.9\linewidth]{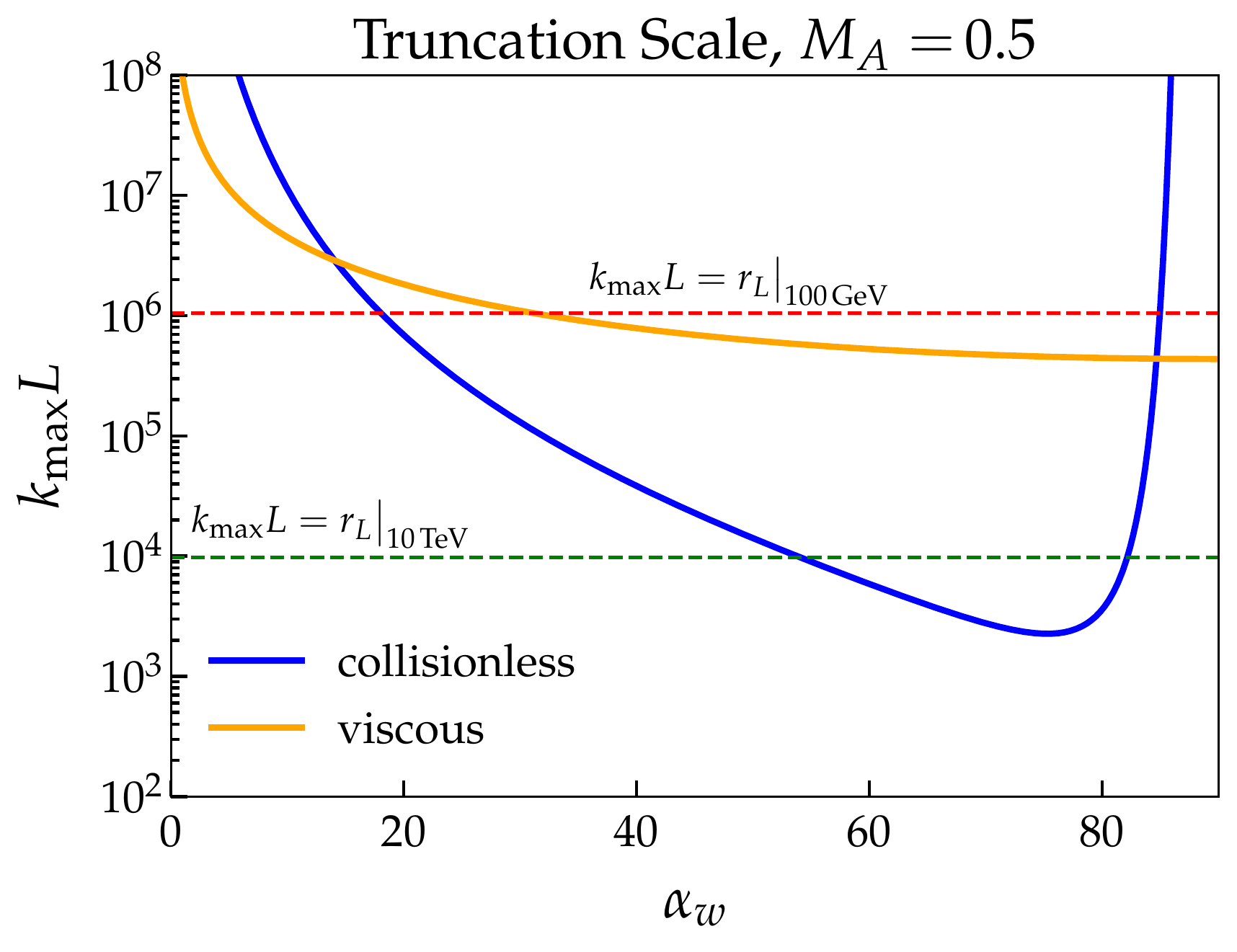}
        \caption{}
    \end{subfigure}
    \hfill
    \begin{subfigure}{0.49\textwidth}
        \centering
        \includegraphics[width=0.9\linewidth]{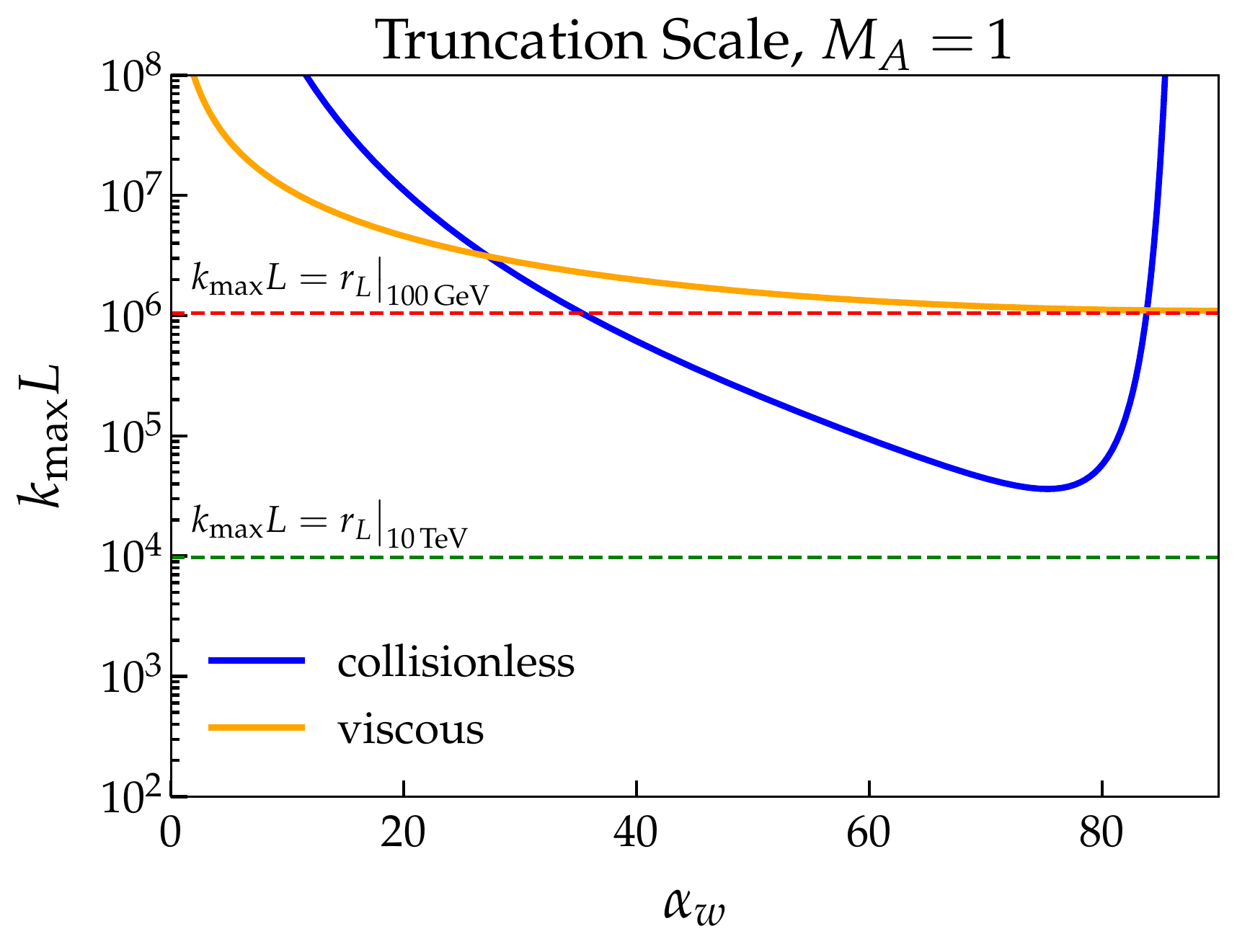}
        \caption{}
    \end{subfigure}
\caption{\small{The truncation scales of the fluctuations as a function of the wave pitch-angle, due to viscous (solid orange lines) and collisionless (solid blue line) dampings are shown, (a) for the case $M_A = 0.5$ and (b) for the case $M_A=1$. The dashed horizontal lines correspond to the values of $k_\mathrm{max} L$ that resonate with Larmor radii of $100 \, \mathrm{GeV}$ particles (dashed red) and of $10 \, \mathrm{TeV}$ particles (dashed green). As an example, it can be seen that, in the case $M_A=1$, $10 \, \mathrm{TeV}$ CRs can scatter against all the waves, since they have not been damped by the medium, while this is not true anymore where $M_A=0.5$, since they can only scatter off waves with a pitch angle up to $\sim 50^\circ$. More scattering implies obviously more confinement, namely a larger $D_{\mu\mu}$ and a smaller (spatial) $D(E)$.}}
\label{fig:truncation_scale}
\end{figure*}

One thing that is worth noticing is the truncation scales resulting from the two damping mechanisms computed with the listed values, that can be seen in Figure \ref{fig:truncation_scale}, where we show them (a) for the case $M_A=0.5$ (Alfv\'en-dominated region in \citet{PhysRevD.106.103015}) and (b) for the case $M_A=1$ (fast-dominated region in \citet{PhysRevD.106.103015}). In the figure we also show (dashed lines) the $k_\mathrm{max} L$ scales (of the cascade) corresponding to fluctuations resonating with $100 \, \mathrm{GeV}$ (red dashed) and $10, \mathrm{TeV}$ (green dashed) CRs, based on their Larmor radii $r_L$, being $E_\mathrm{CR} \in [100 \, \mathrm{GeV}, \, 10 \, \mathrm{TeV}]$ the energy range of the particles considered in our paper. As it can be seen, a little change in the properties of the cascade (its $M_A$) makes, \textit{e.g.}, $10 \, \mathrm{TeV}$ CRs scatter off all the fluctuations of the fast cascade in the $M_A = 1$ case, while less waves are available to scatter off, for the same CRs, in the $M_A=0.5$ case. This difference makes \textit{our} two diffusion coefficients shown in Figure 1a of \citet{PhysRevD.106.103015} to be different by quite large amounts ($\sim 5$ at $E_{\mathrm{CR}} \simeq 10 \, \mathrm{TeV}$ and $\sim 7.5$ at $E_{\mathrm{CR}} \simeq 100 \, \mathrm{GeV}$). 

\section{Effect of a different normalisation on the resulting $\gamma$-ray emission}
As a matter of fact, it has been pointed out in our paper~\citep{PhysRevD.106.103015} (Section IIb, page 5, right column) that the final result is, in terms of the diffusion properties of the CRs, \textit{nearly exclusively} due to the relative normalisations of the diffusion coefficients in the two regions, up to the point that even a different rigidity scaling does not affect it significantly. This is due to the significance maps that we are plotting to show the $\gamma$-ray distribution, that involve relative differences. The referee raised this point and was satisfied with this answer. 

A different matter would be if the energetics of the sources were analyzed in the paper, to study the broken power-law spectrum shown by Fermi-LAT~\citep{Ackermann:2011lfa} + HAWC~\citep{Abeysekara:2021yum}: in that case, a careful evaluation of the age of the sources and the CR injection spectrum would have been required. However, we found it to be pointless, due to the high level of uncertainties involved in those estimations, which would allow us to properly readjust the environmental parameters, still within the reasonable uncertainty band.

\section{Effect of a different injection mechanism on the resulting $\gamma$-ray emission}
For what concerns the injection mechanism, we explicitly describe in the paper that $\sim 90\%$ of the particle emission occurs as an initial burst (Section IIb, page 4, left column). We believe that this is physically motivated by two main reasons:
\begin{itemize}
    \item the acceleration of CRs occurs inside the bubble in a \textit{reverse shock}, so that the region \textit{upstream} of the shock can trap the particles inside the bubble until dissipation of the shock itself, a process that can take as long as $\sim \mathcal{O}(\mathrm{Myr})$~\citep{1:2021xpo}, after which the bulk of the accelerated CRs can be released into the ISM,
    \item there is evidence for a $\sim 1\big/r$ radial profile of the surface brightness detected from the OB cluster~\citep{Aharonian:2018oau}, which should indicate a continuous injection of CR particles from the region during (at least) its most recent stage --- this is a standard calculation: see, \textit{e.g.} \citet{DeLaTorreLuque:2022chz}, Appendix B.
\end{itemize}

We implemented the above considerations as described in the paper. Furthermore, we remark that this is part of the modeling procedure, and as such is considered to obtain the expected outcome.

\section{General discussion}
It is probably worth at this point to clarify what are --- in our opinion --- the key points to learn from any calculation of the diffusion coefficient in these contexts. As shown in Equation \eqref{eq:viscous_parameters}, the calculation of the damping-parameter scales, which is of primary importance in this framework (see \textit{e.g.} Figure \ref{fig:truncation_scale} and its discussion above), is based on the results from previous research. These quantities could be updated when we have a better knowledge of the composition of the background plasma in the ISM. Furthermore, to compute the truncation scales in Equation \eqref{eq:truncation_scale}, we had to estimate the developement timescale of the \textit{fast magnetosonic} cascade, which involves the estimation of the fluid velocity of the turbulence, at each scale --- see, \textit{e.g.}, \citet{Schekochihin:2020aqu} for a recent review on the topic, where the scalings of the \textit{magneto-hydro-dynamic} (MHD) cascades are discussed. Besides, to the best of our knowledge, there is no analytic theory on the MHD \textit{compressible} cascade, but only on the Alfv\'enic one~\citep{Schekochihin:2020aqu}. This quantities, therefore, should be taken with a grain of salt.

With this in mind, what we believe is really relevant in the context of our paper --- and explicitly stated in it --- is the difference in the propagation properties of the CRs between two regions where different modes dominate the diffusion process: (i) where Alfv\'en modes dominate in the background plasma, particles are less trapped and propagate faster, (ii) where fast modes dominate, particles are confined for longer. This has the implications in the $\gamma$-ray emission of the Cygnus-X region that are described in our paper~\citep{PhysRevD.106.103015}. In particular, the diffuse $\gamma$-ray emission in the region is correlated with such difference, rather than with the baryonic-gas distribution. Within reasonable physical uncertainties, all the parameters can be, instead, appropriately tuned. This is the reason why we believe an accurate analysis of the energetics of the sources might not be based on solid grounds.

We remain open to possible quantitative criticism on the results of the paper \citet{PhysRevD.106.103015}, which we certainly have already received during the review process from the Journal. Until then, we sincerely hope that everything is clearer, now.

\newpage
\bibliography{apssamp}

\begin{thebibliography}{16}%
\makeatletter
\providecommand \@ifxundefined [1]{%
 \@ifx{#1\undefined}
}%
\providecommand \@ifnum [1]{%
 \ifnum #1\expandafter \@firstoftwo
 \else \expandafter \@secondoftwo
 \fi
}%
\providecommand \@ifx [1]{%
 \ifx #1\expandafter \@firstoftwo
 \else \expandafter \@secondoftwo
 \fi
}%
\providecommand \natexlab [1]{#1}%
\providecommand \enquote  [1]{``#1''}%
\providecommand \bibnamefont  [1]{#1}%
\providecommand \bibfnamefont [1]{#1}%
\providecommand \citenamefont [1]{#1}%
\providecommand \href@noop [0]{\@secondoftwo}%
\providecommand \href [0]{\begingroup \@sanitize@url \@href}%
\providecommand \@href[1]{\@@startlink{#1}\@@href}%
\providecommand \@@href[1]{\endgroup#1\@@endlink}%
\providecommand \@sanitize@url [0]{\catcode `\\12\catcode `\$12\catcode
  `\&12\catcode `\#12\catcode `\^12\catcode `\_12\catcode `\%12\relax}%
\providecommand \@@startlink[1]{}%
\providecommand \@@endlink[0]{}%
\providecommand \url  [0]{\begingroup\@sanitize@url \@url }%
\providecommand \@url [1]{\endgroup\@href {#1}{\urlprefix }}%
\providecommand \urlprefix  [0]{URL }%
\providecommand \Eprint [0]{\href }%
\providecommand \doibase [0]{http://dx.doi.org/}%
\providecommand \selectlanguage [0]{\@gobble}%
\providecommand \bibinfo  [0]{\@secondoftwo}%
\providecommand \bibfield  [0]{\@secondoftwo}%
\providecommand \translation [1]{[#1]}%
\providecommand \BibitemOpen [0]{}%
\providecommand \bibitemStop [0]{}%
\providecommand \bibitemNoStop [0]{.\EOS\space}%
\providecommand \EOS [0]{\spacefactor3000\relax}%
\providecommand \BibitemShut  [1]{\csname bibitem#1\endcsname}%
\let\auto@bib@innerbib\@empty
\bibitem [{\citenamefont {Yan}\ and\ \citenamefont
  {Pavaskar}(2022)}]{Yan:2022mma}%
  \BibitemOpen
  \bibfield  {author} {\bibinfo {author} {\bibfnamefont {H.}~\bibnamefont
  {Yan}}\ and\ \bibinfo {author} {\bibfnamefont {P.}~\bibnamefont {Pavaskar}},\
  }\href@noop {} {\  (\bibinfo {year} {2022})},\ \Eprint
  {http://arxiv.org/abs/2211.17057} {arXiv:2211.17057 [astro-ph.HE]}
  \BibitemShut {NoStop}%
\bibitem [{\citenamefont {Fornieri}\ and\ \citenamefont
  {Zhang}(2022)}]{PhysRevD.106.103015}%
  \BibitemOpen
  \bibfield  {author} {\bibinfo {author} {\bibfnamefont {O.}~\bibnamefont
  {Fornieri}}\ and\ \bibinfo {author} {\bibfnamefont {H.}~\bibnamefont
  {Zhang}},\ }\href {\doibase 10.1103/PhysRevD.106.103015} {\bibfield
  {journal} {\bibinfo  {journal} {Phys. Rev. D}\ }\textbf {\bibinfo {volume}
  {106}},\ \bibinfo {pages} {103015} (\bibinfo {year} {2022})}\BibitemShut
  {NoStop}%
\bibitem [{\citenamefont {{Kulsrud}}\ and\ \citenamefont
  {{Pearce}}(1969)}]{1969ApJ...156..445K}%
  \BibitemOpen
  \bibfield  {author} {\bibinfo {author} {\bibfnamefont {R.}~\bibnamefont
  {{Kulsrud}}}\ and\ \bibinfo {author} {\bibfnamefont {W.~P.}\ \bibnamefont
  {{Pearce}}},\ }\href {\doibase 10.1086/149981} {\bibfield  {journal}
  {\bibinfo  {journal} {\apj}\ }\textbf {\bibinfo {volume} {156}},\ \bibinfo
  {pages} {445} (\bibinfo {year} {1969})}\BibitemShut {NoStop}%
\bibitem [{\citenamefont {{Voelk}}(1975)}]{1975RvGSP..13..547V}%
  \BibitemOpen
  \bibfield  {author} {\bibinfo {author} {\bibfnamefont {H.~J.}\ \bibnamefont
  {{Voelk}}},\ }\href {\doibase 10.1029/RG013i004p00547} {\bibfield  {journal}
  {\bibinfo  {journal} {Reviews of Geophysics and Space Physics}\ }\textbf
  {\bibinfo {volume} {13}},\ \bibinfo {pages} {547} (\bibinfo {year}
  {1975})}\BibitemShut {NoStop}%
\bibitem [{\citenamefont {Fornieri}\ \emph {et~al.}(2021)\citenamefont
  {Fornieri}, \citenamefont {Gaggero}, \citenamefont {Cerri}, \citenamefont
  {De~la Torre~Luque},\ and\ \citenamefont {Gabici}}]{Fornieri:2020wrr}%
  \BibitemOpen
  \bibfield  {author} {\bibinfo {author} {\bibfnamefont {O.}~\bibnamefont
  {Fornieri}}, \bibinfo {author} {\bibfnamefont {D.}~\bibnamefont {Gaggero}},
  \bibinfo {author} {\bibfnamefont {S.~S.}\ \bibnamefont {Cerri}}, \bibinfo
  {author} {\bibfnamefont {P.}~\bibnamefont {De~la Torre~Luque}}, \ and\
  \bibinfo {author} {\bibfnamefont {S.}~\bibnamefont {Gabici}},\ }\href
  {\doibase 10.1093/mnras/stab355} {\bibfield  {journal} {\bibinfo  {journal}
  {Mon. Not. Roy. Astron. Soc.}\ }\textbf {\bibinfo {volume} {502}},\ \bibinfo
  {pages} {5821} (\bibinfo {year} {2021})},\ \Eprint
  {http://arxiv.org/abs/2011.09197} {arXiv:2011.09197 [astro-ph.HE]}
  \BibitemShut {NoStop}%
\bibitem [{\citenamefont {Ginzburg}\ \emph {et~al.}(1970)\citenamefont
  {Ginzburg}, \citenamefont {Ginzburg}, \citenamefont {Sykes},\ and\
  \citenamefont {Tayler}}]{ginzburg1970propagation}%
  \BibitemOpen
  \bibfield  {author} {\bibinfo {author} {\bibfnamefont {V.}~\bibnamefont
  {Ginzburg}}, \bibinfo {author} {\bibfnamefont {V.}~\bibnamefont {Ginzburg}},
  \bibinfo {author} {\bibfnamefont {J.}~\bibnamefont {Sykes}}, \ and\ \bibinfo
  {author} {\bibfnamefont {R.}~\bibnamefont {Tayler}},\ }\href
  {https://books.google.it/books?id=vF55AAAAIAAJ} {\emph {\bibinfo {title} {The
  Propagation of Electromagnetic Waves in Plasmas}}},\ Commonwealth and
  International Library\ (\bibinfo  {publisher} {Pergamon Press},\ \bibinfo
  {year} {1970})\BibitemShut {NoStop}%
\bibitem [{\citenamefont {{Braginskii}}(1965)}]{1965RvPP....1..205B}%
  \BibitemOpen
  \bibfield  {author} {\bibinfo {author} {\bibfnamefont {S.~I.}\ \bibnamefont
  {{Braginskii}}},\ }\href@noop {} {\bibfield  {journal} {\bibinfo  {journal}
  {Reviews of Plasma Physics}\ }\textbf {\bibinfo {volume} {1}},\ \bibinfo
  {pages} {205} (\bibinfo {year} {1965})}\BibitemShut {NoStop}%
\bibitem [{\citenamefont {{Petrosian}}\ \emph {et~al.}(2006)\citenamefont
  {{Petrosian}}, \citenamefont {{Yan}},\ and\ \citenamefont
  {{Lazarian}}}]{2006ApJ...644..603P}%
  \BibitemOpen
  \bibfield  {author} {\bibinfo {author} {\bibfnamefont {V.}~\bibnamefont
  {{Petrosian}}}, \bibinfo {author} {\bibfnamefont {H.}~\bibnamefont {{Yan}}},
  \ and\ \bibinfo {author} {\bibfnamefont {A.}~\bibnamefont {{Lazarian}}},\
  }\href {\doibase 10.1086/503378} {\bibfield  {journal} {\bibinfo  {journal}
  {\apj}\ }\textbf {\bibinfo {volume} {644}},\ \bibinfo {pages} {603} (\bibinfo
  {year} {2006})},\ \Eprint {http://arxiv.org/abs/astro-ph/0508567}
  {arXiv:astro-ph/0508567 [astro-ph]} \BibitemShut {NoStop}%
\bibitem [{\citenamefont {Yan}\ and\ \citenamefont
  {Lazarian}(2008)}]{Yan:2007uc}%
  \BibitemOpen
  \bibfield  {author} {\bibinfo {author} {\bibfnamefont {H.}~\bibnamefont
  {Yan}}\ and\ \bibinfo {author} {\bibfnamefont {A.}~\bibnamefont {Lazarian}},\
  }\href {\doibase 10.1086/524771} {\bibfield  {journal} {\bibinfo  {journal}
  {Astrophys. J.}\ }\textbf {\bibinfo {volume} {673}},\ \bibinfo {pages} {942}
  (\bibinfo {year} {2008})},\ \Eprint {http://arxiv.org/abs/0710.2617}
  {arXiv:0710.2617 [astro-ph]} \BibitemShut {NoStop}%
\bibitem [{\citenamefont {{Spitzer}}(1962)}]{1962pfig.book.....S}%
  \BibitemOpen
  \bibfield  {author} {\bibinfo {author} {\bibfnamefont {L.}~\bibnamefont
  {{Spitzer}}},\ }\href@noop {} {\emph {\bibinfo {title} {{Physics of Fully
  Ionized Gases}}}}\ (\bibinfo {year} {1962})\BibitemShut {NoStop}%
\bibitem [{\citenamefont {Ackermann}\ \emph {et~al.}(2011)\citenamefont
  {Ackermann} \emph {et~al.}}]{Ackermann:2011lfa}%
  \BibitemOpen
  \bibfield  {author} {\bibinfo {author} {\bibfnamefont {M.}~\bibnamefont
  {Ackermann}} \emph {et~al.},\ }\href {\doibase 10.1126/science.1210311}
  {\bibfield  {journal} {\bibinfo  {journal} {Science}\ }\textbf {\bibinfo
  {volume} {334}},\ \bibinfo {pages} {1103} (\bibinfo {year}
  {2011})}\BibitemShut {NoStop}%
\bibitem [{\citenamefont {Abeysekara}\ \emph {et~al.}(2021)\citenamefont
  {Abeysekara} \emph {et~al.}}]{Abeysekara:2021yum}%
  \BibitemOpen
  \bibfield  {author} {\bibinfo {author} {\bibfnamefont {A.~U.}\ \bibnamefont
  {Abeysekara}} \emph {et~al.},\ }\href {\doibase 10.1038/s41550-021-01318-y}
  {\bibfield  {journal} {\bibinfo  {journal} {Nature Astron.}\ }\textbf
  {\bibinfo {volume} {5}},\ \bibinfo {pages} {465} (\bibinfo {year} {2021})},\
  \Eprint {http://arxiv.org/abs/2103.06820} {arXiv:2103.06820 [astro-ph.HE]}
  \BibitemShut {NoStop}%
\bibitem [{\citenamefont {1}\ \emph {et~al.}(2021)\citenamefont {1},
  \citenamefont {Blasi}, \citenamefont {Peretti},\ and\ \citenamefont
  {Cristofari}}]{1:2021xpo}%
  \BibitemOpen
  \bibfield  {author} {\bibinfo {author} {\bibfnamefont {G.~M.}\ \bibnamefont
  {1}}, \bibinfo {author} {\bibfnamefont {P.}~\bibnamefont {Blasi}}, \bibinfo
  {author} {\bibfnamefont {E.}~\bibnamefont {Peretti}}, \ and\ \bibinfo
  {author} {\bibfnamefont {P.}~\bibnamefont {Cristofari}},\ }\href {\doibase
  10.1093/mnras/stab690} {\  (\bibinfo {year} {2021}),\
  10.1093/mnras/stab690},\ \Eprint {http://arxiv.org/abs/2102.09217}
  {arXiv:2102.09217 [astro-ph.HE]} \BibitemShut {NoStop}%
\bibitem [{\citenamefont {Aharonian}\ \emph {et~al.}(2019)\citenamefont
  {Aharonian}, \citenamefont {Yang},\ and\ \citenamefont {de~O\~na
  Wilhelmi}}]{Aharonian:2018oau}%
  \BibitemOpen
  \bibfield  {author} {\bibinfo {author} {\bibfnamefont {F.}~\bibnamefont
  {Aharonian}}, \bibinfo {author} {\bibfnamefont {R.}~\bibnamefont {Yang}}, \
  and\ \bibinfo {author} {\bibfnamefont {E.}~\bibnamefont {de~O\~na
  Wilhelmi}},\ }\href {\doibase 10.1038/s41550-019-0724-0} {\bibfield
  {journal} {\bibinfo  {journal} {Nature Astron.}\ }\textbf {\bibinfo {volume}
  {3}},\ \bibinfo {pages} {561} (\bibinfo {year} {2019})},\ \Eprint
  {http://arxiv.org/abs/1804.02331} {arXiv:1804.02331 [astro-ph.HE]}
  \BibitemShut {NoStop}%
\bibitem [{\citenamefont {De~La Torre~Luque}\ \emph {et~al.}(2022)\citenamefont
  {De~La Torre~Luque}, \citenamefont {Fornieri},\ and\ \citenamefont
  {Linden}}]{DeLaTorreLuque:2022chz}%
  \BibitemOpen
  \bibfield  {author} {\bibinfo {author} {\bibfnamefont {P.}~\bibnamefont
  {De~La Torre~Luque}}, \bibinfo {author} {\bibfnamefont {O.}~\bibnamefont
  {Fornieri}}, \ and\ \bibinfo {author} {\bibfnamefont {T.}~\bibnamefont
  {Linden}},\ }\href@noop {} {\  (\bibinfo {year} {2022})},\ \Eprint
  {http://arxiv.org/abs/2205.08544} {arXiv:2205.08544 [astro-ph.HE]}
  \BibitemShut {NoStop}%
\bibitem [{\citenamefont {Schekochihin}(2020)}]{Schekochihin:2020aqu}%
  \BibitemOpen
  \bibfield  {author} {\bibinfo {author} {\bibfnamefont {A.~A.}\ \bibnamefont
  {Schekochihin}},\ }\href@noop {} {\  (\bibinfo {year} {2020})},\ \Eprint
  {http://arxiv.org/abs/2010.00699} {arXiv:2010.00699 [physics.plasm-ph]}
  \BibitemShut {NoStop}%
\end{thebibliography}%
\bibliographystyle{apsrev4-1}
\end{document}